\documentclass{elsart3-1}

 \usepackage{graphicx}

\usepackage{amssymb}

\usepackage[english,francais]{babel}

\newtheorem{e-proposition}[theorem]{Proposition}

\newtheorem{e-definition}[theorem]{Definition\rm}

\renewcommand{\theequation}{\arabic{equation}}
\def\og{\leavevmode\raise.3ex\hbox{$\scriptscriptstyle\langle\!\langle$~}}
\def\fg{\leavevmode\raise.3ex\hbox{~$\!\scriptscriptstyle\,\rangle\!\rangle$}}

\usepackage{txfonts}
\def\tens#1{\ensuremath{\mathsf{#1}}}

\def\MS#1#2#3{{\tens{M^{#2}_{#1}}#3}}

\begin{document}
\centerline{The next generation radiotelescopes/ Les radiot\'elescopes du 
futur}
\begin{frontmatter}


\selectlanguage{english}

\title{High Dynamic-Range Radio-Interferometric Images at 327 MHz}


\selectlanguage{english}
\author[authorlabel1]{Juan M. Uson},
\ead {juan.uson AT obspm.fr}
\author[authorlabel2]{William D. Cotton}
\ead{bcotton AT nrao.edu}

\address[authorlabel1]{Observatoire de Paris - LERMA, 61 Avenue de l'Observatoire, F75014 Paris, France}
\address[authorlabel2]{NRAO, 520 Edgemont Road, Charlottesville, VA 22903, USA}

\begin{abstract}
Radio astronomical imaging using aperture synthesis telescopes requires deconvolution of the point spread function as well as calibration of the instrumental characteristics (primary beam) and foreground (ionospheric/atmospheric) effects.  These effects vary in time and also across the field of view, resulting in directionally-dependent (DD), time-varying gains.  The primary beam will deviate from the theoretical estimate in real cases at levels that will limit the dynamic range of images if left uncorrected.  Ionospheric electron density variations cause time and position variable refraction of sources.  At low frequencies and sufficiently high dynamic range this will also defocus the images producing error patterns that vary with position and also with frequency due to the chromatic aberration of synthesis telescopes.  Superposition of such residual sidelobes can lead to spurious spectral signals.  Field-based ionospheric calibration as well as ``peeling'' calibration of strong sources leads to images with higher dynamic range and lower spurious signals but will be limited by sensitivity on the necessary short-time scales.  The results are improved images although some artifacts remain.

{\it To cite this article: J. M. Uson, W. D. Cotton, C. R. Physique 13 (2012).}

\vskip 0.5\baselineskip

\selectlanguage{francais}
\noindent{\bf R\'esum\'e}
\vskip 0.5\baselineskip
\noindent
{\bf Synth\`ese d'Images Radio-Interf\'erom\'etriques \`a Haute Dynamique \`a 327 MHz.}

\noindent La formation des images radio-astronomiques obtenues avec un r\'eseau de t\'elescopes de synth\`ese n\'ecessite la d\'econvolution de la fonction de r\'eponse \`a une source ponctuelle ainsi que l'\'etalonnage des caract\'eristiques instrumentales (faisceau primaire) et des effets de propagation (ionosph\`ere/atmosph\`ere). Ces effets varient avec la position et aussi avec le temps. Le faisceau primaire r\'eel va d\'evier de l'estimation th\'eorique \`a des niveaux qui vont limiter la dynamique des images \`a moins qu'on ne les corrige.  Les variations de densit\'e d'\'electrons ionosph\'eriques vont provoquer une r\'efraction variable d\'ependant de la position des sources et du temps. Aux basses fr\'equences et \`a suffisamment haute dynamique les images seront d\'efocalis\'ees de fa\c con variable d\'ependant de la position et aussi de la fr\'equence, \`a cause de l'aberration chromatique des t\'elescopes de synth\`ese.  La superposition des lobes r\'esiduels peut alors g\'en\'erer de faux signaux spectraux. L'\'etalonnage de la r\'efraction ionosph\'erique en fonction de la position et du temps ainsi que l'autocalibration sur les sources les plus puissantes (``peeling'') permettent d'augmenter la dynamique des images et de diminuer les signaux parasites, mais ils sont limit\'es par la sensibilit\'e disponible aux courts temps de pose.  On obtient ainsi des images am\'elior\'ees,  bien que certains artefacts demeurent.

{\it Pour citer cet article~: J. M. Uson, W. D. Cotton, C. R. Physique 13 (2012).}

 \vskip 0.5\baselineskip
\noindent{\small{\it Mots-cl\'es~:} techniques: traitement des images --- techniques: interf\'erom\'etrie --- m\'ethodes : analyse des donn\'ees}}

\end{abstract}

\end{frontmatter}

\selectlanguage{english}
\section{Introduction}
\label{intro}
Radio interferometers are linear devices that evaluate the correlations of electromagnetic radiation received at its component antennas and yield, to first order, estimates of the Fourier transform of the sky distribution [1].  Imaging consists of estimating the true sky brightness from the observed visibilities, a non-linear process which can be divided in three steps:
(1)	ÔRawÕ imaging, simply the Fourier inversion of the visibilities, with weighting used to modify the point-spread function and noise characteristics to control the resulting sidelobe pattern
(2)	Deconvolution  to correct for ÔmissedÕ visibilities which is a non-linear process and where different methods lead to somewhat different results, and 
(3)	Self-calibration where the visibilities are corrected to sharpen the image by improving the calibration, which is also non-linear and requires significant sensitivity.

Actual interferometers are imperfect devices and suffer from variable foreground effects.  In practice, interferometers do not simply measure the Fourier transform of the sky distribution but their measurements must be described by their so-called Ômeasurement equationÕ as indicated in the following (matrix) equation [2]:

\begin{equation}
\label{ME}
V^{Obs}_{ij} = \MS{ij}{}{}\int \MS{ij}{Sky}{}(\mathbf{s}) I(\mathbf{s}) e^{2\pi\iota \mathbf{s} \cdot \mathbf{b_{ij}}} d \mathbf{s}
\end{equation}

where $I(\mathbf{s})$ describes the flux-density distribution on the sky, $V^{Obs}_{ij}$ is the full-polarization visibility observed by baseline ij,  \MS{ij}{}\  are the Mueller matrices that describe the directionally-independent gains, and \MS{ij}{Sky}{}$(\mathbf{s})$ are the Mueller matrices that describe directionally-dependent gains.  Generally, both sets of gain matrices vary with time.

The directionally-independent effects describe for example the average calibration terms for each antenna at any given time as well as the average primary-beam response which conventionally is taken as an azimuthally-symmetric modulation of the sky distribution (and is simply inversely-applied to the final image) whilst the directionally-dependent effects describe most propagation effects, non-isoplanatic calibration terms as well as departures from the ideal primary-beam response [3].  To first order, the average primary-beam response can be incorporated into the ``observed'' flux distribution on the sky and the directionally-dependent calibration terms can be ignored which leads to the familiar, first-order Fourier relation between the observed sky distribution and the measured visibilities.  However, if such directionally-dependent corrections can be evaluated, they can be introduced easily in the ``reverse'' step of the imaging cycle pioneered by Cotton and Schwab [4,5], allowing a progressive improvement to the dynamic range of the images thus obtained.

Electromagnetic signals propagating through an ionized medium experience an excess phase delay inversely-proportional to the frequency-squared which can be a serious source of phase corruptions for radiation at low frequencies.  The Earth's ionosphere is such an ionized medium and its irregularities have long been recognized as the major source of position and time-dependent phase corruption for high resolution arrays at frequencies below ~100 MHz.  However, because the resolution of an array is proportional to the observation frequency, the effects of ionospheric phase corruption on derived images vary as the inverse of the frequency.  Because of the large field of view at the low frequencies at which the ionospheric effects are most relevant, these effects appear as directionally-dependent distortions of the type described by  \MS{ij}{Sky}{}$(\mathbf{s})$ in equation~(\ref{ME}) above.

As is well known, for sufficiently small fields of view directionally-dependent effects can be ignored and the measurement equation can be written as a 2D Fourier transform

\begin{equation}
\label{2DME}
V^{*}_{ij} (u,v) = {}{}\int  I(l,m) e^{2\pi\iota (u l + v m)} dl dm
\end{equation}

Inversion of this equation is standard practice and can be done using 2D FFTs supported on facets tangent to the celestial sphere or their projections onto a 2D plane tangent to the celestial sphere at a given position, usually the pointing center [6].  As indicated above, Cotton-Schwab deconvolution can be used to lessen the required dynamic range progressively by introducing the necessary directionally-dependent factors as deconvolution proceeds, which approaches inversion of equation~(\ref{ME}) with steadily increasing precision as far as knowledge of the directionally-dependent terms permits.  A variation of this procedure allows the introduction of time-variable, directionally-dependent corrections (calibrations) to the first step in this procedure in which the data are gridded and the first-order images are determined, as discussed in section~\ref{fcal}.  Some of the necessary directionally-dependent corrections are a-priori unknown, such as departures of the primary beam response from cylindrical symmetry, but can limit the dynamic range of the images.  We show in section~\ref{peel} that limited ``peeling'' can be used to improve dynamic range by determining these corrections from the data but precision is limited by sensitivity and degrees of freedom.  All algorithms discussed in this paper can be found in the ``Obit'' package [7].

\section{Observations}
\label{obs}

We have studied the effects of ionospheric perturbations on data obtained at $\sim 327$~MHz applying the techniques developed for lower frequency observations [8,9].  We have applied the field-based calibration technique to a deep observation of an ``empty'' field with the VLA\footnote{The Very Large Array (VLA) of the National Radio Astronomy Observatory is a facility of the National Science Foundation, operated under cooperative agreement by Associated Universities, Inc.} in its most extended (A) configuration using two optimal spectral windows (315-318 MHz and 325-328 MHz), chosen to avoid contamination by strong ``5 MHz birdies,'' internally-generated interference signals which were effectively aliased to DC at the low-frequency end of each spectrum and cut-off by the baseband filter roll-offs at the other end.

After standard calibration using 3C286 (observed for 4 minutes in each 15 minute observing cycle) for phase, amplitude and bandpass determination, images were made and the data self-calibrated using standard (3D) algorithms.  Data contaminated by interference (hereafter RFI) where identified using a variety of algorithms with excision of low-level contamination performed using the ``UVLIN'' algorithm [10] in which a first-order polynomial is fit to each visibility as a function of frequency (real and imaginary components fit separately) and the visibility is discarded if any of the residuals exceeds a threshold of $8 \sigma$, supplemented by discarding observations for which the self-calibration did not converge.  All channels of a contaminated visibility sample were discarded in order to avoid the need for introducing ``effective'' corrections to the average imaged frequency which cannot be handled at this time by the algorithms that we use.  This is a small, but subtle, systematic effect which is usually ignored but which might become important for observations with high dynamic range such as those envisioned for the Square Kilometre Array (SKA).

\begin{figure}
   \centering
   \includegraphics[height=3.4in]{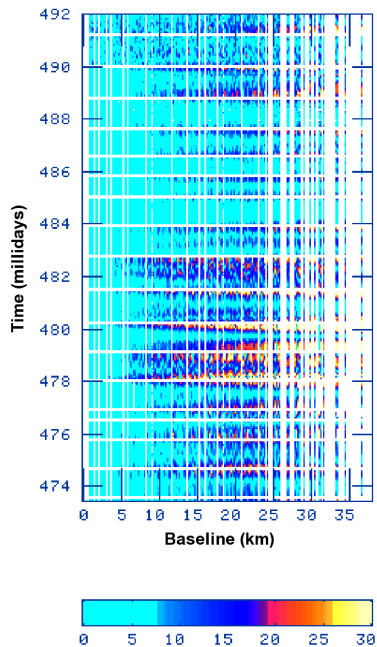}
\caption{ 
``Waterfall'' plot of the (square-root of the) ionospheric phase-structure function derived from observations of the calibrator, 3C286.  Some intervals show little ionospheric activity while disturbances are quite noticeable at other times. Each ``scan'' has been calibrated previously.  The horizontal gaps correspond to time intervals of $\sim$11~minutes when the target field was observed.  The color scale is labeled in degrees.
}
\label{water}
\end{figure}

\section{Field-based calibrations}
\label{fcal}

An electromagnetic wavefront passing through a medium with position-dependent index of refraction will experience variable phase delays and attenuation in different parts of the wavefront.  A wedge in the density of such a medium will cause (to first order) a linear phase gradient across an array observing through said medium resulting in refraction of any sources seen.  Higher-order variations in the index of refraction will cause more substantial distortions to the wavefront producing defocusing and scintillation in extreme cases as well as attenuation (ignored in this discussion).  Calibration of the directionally-dependent terms in the measurement equation involves estimating and correcting these phase corruptions.

Ionospheric effects can be characterized and visualized in a variety of ways.  One of these is the structure function

\begin{equation}
\label{SF}
\it{SF}(x) = {}{}\langle ( \phi (x_0) - \phi (x_0 + x) )^2 \rangle
\end{equation}

where $x$ and $x_0$ are positions of antennas of the array.  The square-root of the {\it SF} is the (rms) phase difference as a function of the separation of any two antennas.  Figure~\ref{water} shows a ``waterfall'' plot of this quantity as a function of baseline and time with time increasing upwards.  Notice that the baseline lengths neglect projection effects and that the gaps in time have been eliminated as
each block corresponds to one of the calibrator scans and the 11-minute periods of observations of the target field have been compressed.

Because of the wide field-of-view (FOV) that the VLA antennas see at this frequency [11,12], it is necessary to determine position-dependent phase corrections even though this provides only a first-order correction as position-dependent absorption is neglected at this stage.  In the regime in which the phase screen can be described adequately as a linear gradient across the array for a wavefront coming from any given direction, the image of a small source will not be distorted but only shifted from its true position.  In this regime, the ``field-based'' calibration technique [4,5] is applicable.  The method consists of using snapshot measurements of small fields centered on the positions of known, moderately strong sources in the FOV, deconvolve such images and determine the offsets in their apparent positions for each time interval.  This leads to a time series of geometric distortions of the sky as seen by the array.  Low-order Zernike polynomials (2nd order, 5 terms for the example shown here) are used to model the distortion field at each time interval.  The field is modeled as a phase screen and each determined position offset yields a 2-D gradient in the screen at the position given by the line-of-sight to the calibrator.  There is no simple operation that can be applied to the data to introduce these calibrations due to their directional dependence but they can be applied in the course of the imaging and deconvolution operations as indicated above.  Given the large FOV as well as higher order distortions to the wavefront, higher-order phase terms might be necessary for a full description which would necessitate using higher order Zernike polynomials (or other suitable base) to describe and determine the phase screen.

Ionospheric phase corruptions at 327 MHz at the VLA are mostly mild which allows for conventional self-calibration to yield an adequate first-order image of the observed field.  In addition, the galactic background emission does not overwhelm the emission of point sources at 327 MHz (which allows for good sensitivity to detect them) and the ionospheric coherence time is also sufficiently long to allow the detection of about 50 sources even in short (few-minute) snapshots, thus allowing, in principle, a more complex model than a 5-term Zernike description of the ionosphere as a wedge as must be done at the lowest VLA frequency of 74 MHz.  Nevertheless, the example presented here has been processed with the simpler 5-term Zernike model (see Fig. 2).

Snapshot images were made over a FOV with radius of $10^{\rm o}$ with suitable ``flanking fields'' to support the brightest sources.  We have used a catalog of such sources derived from the NVSS [13] with improved positions obtained from the observations discussed here.  The strongest 40~sources (as actually determined by these observations) have been imaged for each snapshot.  However, only those showing a well-defined core have been used to determine the Zernike coefficients although all of them have been included in the snapshot images to minimize artifacts from non-deconvolved sources.  This process results in a calibrator catalog appropriate to these observations which contains all bright sources with an indication of the usability of each entry, i.e. an indication of crowding and structure.  The 2-minute snapshots are used to determine the ionospheric Zernike polynomials which are stored as a table to be used in subsequent imaging.

\begin{figure}
   \centering
   \centerline{
   \includegraphics[height=3.0in]{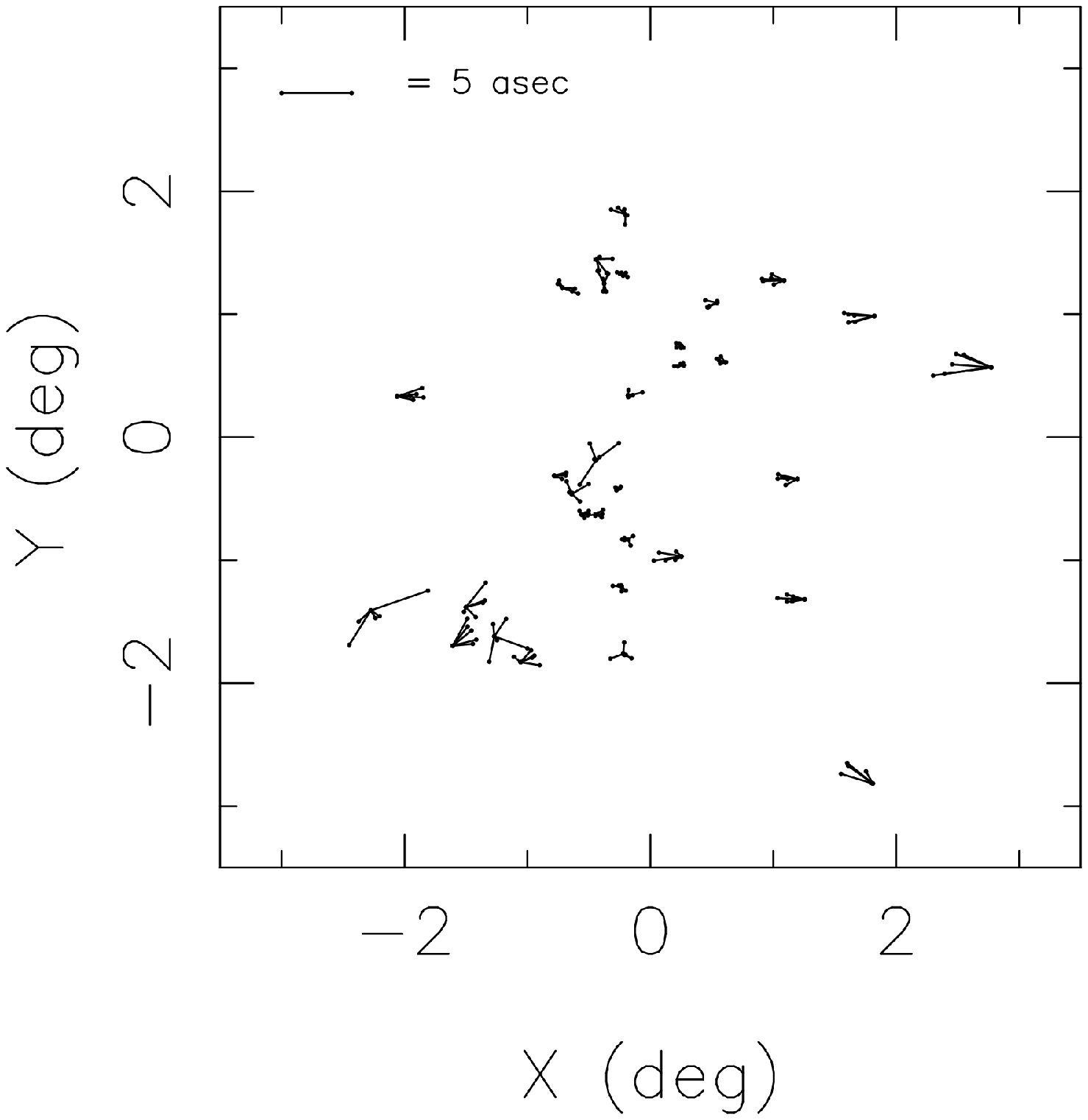}
   \hfill
   \includegraphics[height=3.0in]{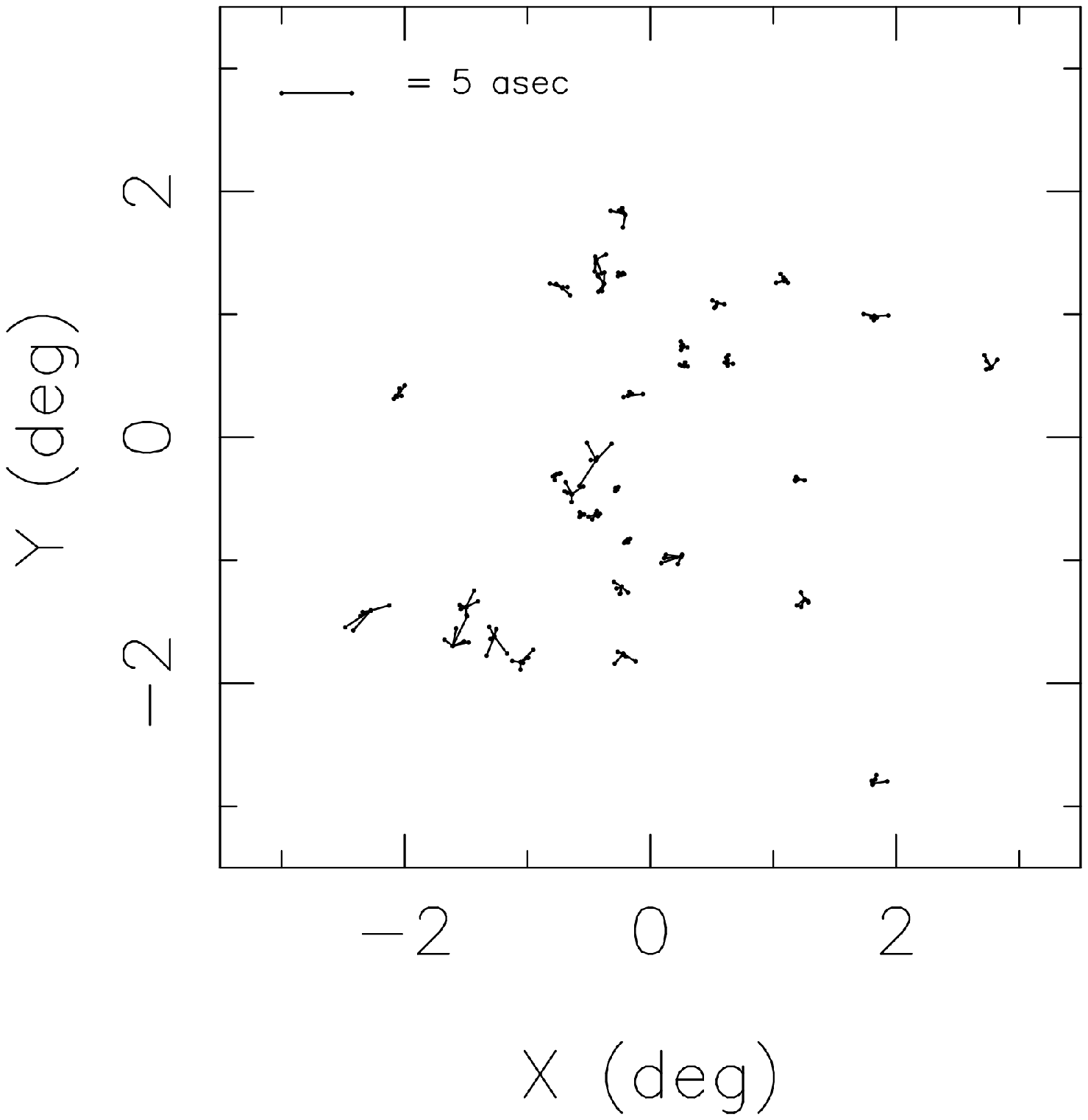}
}

\caption{ 
The figure shows the positions of sources used to determine six of the ionospheric phase screens (here in 3-minute intervals for one of the two sub-bands) midway through the observations with the position offsets multiplied by 500.  Lines are drawn between the actual and apparent position for each solution.  On the left are the observed offsets, on the right the residuals after correction.  A ``scaled''  $5''$ scale-bar is given on the upper left corner of each plot.  The residuals increase near the fiducial ``nulls'' of the primary-beam response.
}
\label{residuals}
\end{figure}

\begin{figure}
   \centering
   \includegraphics[height=3.5in]{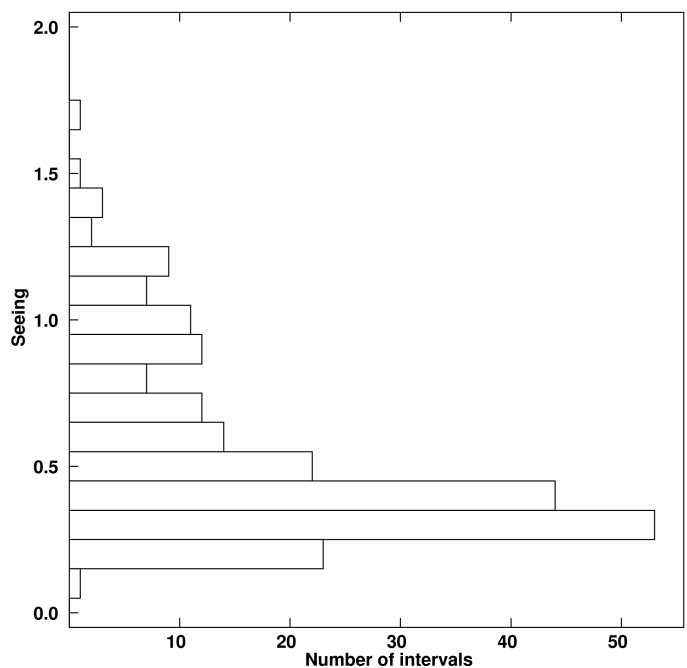}
\caption{ 
The figure shows the residual rms errors to the ionospheric phase screen fits (2-minute intervals, both 3~MHz sub-bands included).  The peak corresponds to the ``quiet'' periods and reflects the rms sensitivity obtained in every 2-minute interval.  The tail corresponds to the non-fitted higher-order perturbations of the ionosphere.  A cutoff of $1.2''$ (20\% of the resolution) leads to discarding 3\% of the observations.
}
\label{seeing}
\end{figure}

For our data, the field-based calibration yields a best-fit solution for each 2-minute time interval when both sub-bands are combined, as well as an rms residual (seeing) error.  Because the VLA is $\sim$\ $0.4 \lambda$ out-of-focus due to the limitation posed by the feed legs that support the prime-focus feed and because the subreflector (which acts as a ground plane) is tilted, the beam lacks true nulls and at the regions where the sensitivity drops in search of such fiducial nulls, there are necessarily steep phase gradients [11,12].  This leads to non-azimuthally-symmetric  primary beams and imposes significant variability to sources observed in these regions.  After dropping some of such sources that might appear too weak in a given 2-minute solution, we arrive at a final solution and associated rms for the ionospheric screen in each 2-minute interval (figure~\ref{seeing}).  We impose a $1.2''$ cutoff, appropriate to our $\sim$\ $6''$ synthesized beam, which results in discarding about 3\% of the data.  Images are subsequently made that also include auto-centering of strong sources to avoid dynamic-range limiting deconvolution errors and also to accelerate convergence [14].

\section{Imaging}
\label{imag}

It is well known that imaging fields with moderate to large FOV needs to address the so-called ``W'' problem, that images are flat but the sky is not [15].  Equation~\ref{2DME} can be applied to each facet to generate a set of first-order images which will be dynamic-range limited, but this limit will be progressively lessened as ``Cotton-Schwab'' deconvolution progresses while allowing the incorporation of directionally-dependent corrections as indicated above.  A number of solutions address the W-problem but will not be discussed here.  Obit allows a 3-D tiling of the FOV with (optionally) subsequent projection onto a single 2D plane typically tangent to the celestial sphere at the pointing center.  Minor shifting of the 3D tiles in order to incorporate all projections onto a common grid [6], can greatly speed-up deconvolution without the memory requirements of other solutions while keeping the necessary tangent sub-fields at the positions of the strongest sources [13,14].

Images were made using a ``fly's eye'' mosaic covering a FOV with radius of $2^{\rm o}$ (511~fields), adding suitable ``flanking fields'' (9~fields) to support sources within a radius of $10^{\rm o}$ whose strength overcomes the attenuation of the primary beam even outside the main lobe.  Initially, conventional imaging was used to self-calibrate the data using a 1-minute interval for phase self-calibration (2 iterations) and a 30-minute interval for subsequent amplitude and phase self-calibration.  Auto-centering of sources with peak brightness higher than 50~mJy/beam was used in all imaging computations [14].  The ionospheric corrections are used in two different steps, first to determine phase corrections to the center of each facet ($768'' \times 768''$, with an effective, undistorted field-of-view of $728''$ in diameter) and auxiliary field which are used in gridding of each field with the 2-minute phase corrections applied on the fly and, second, to correct the phase of each clean component as appropriate in the ``Cotton-Schwab'' subtraction from the visibilities.  Sources observed at strenghts above 50~mJy/beam are imaged at the centers of auxiliary facets ($128'' \times 128''$) in order to minimize residual errors in their vicinity [14] The ionospheric corrections lead to images with higher dynamic range (see Fig. 4).
\begin{figure*}
\centering
\centerline{
   \includegraphics[height=2.5in]{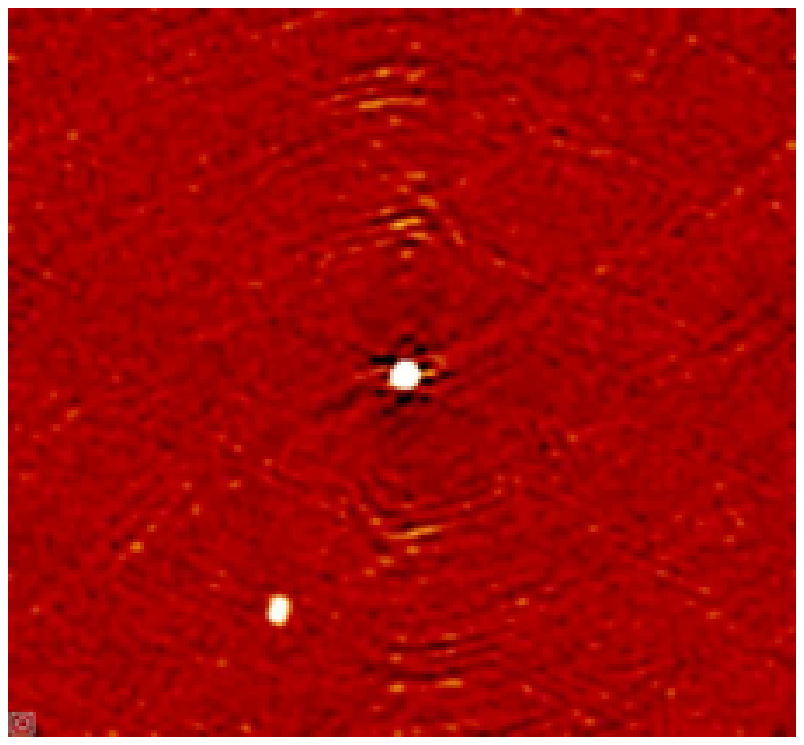}
   \includegraphics[height=2.5in]{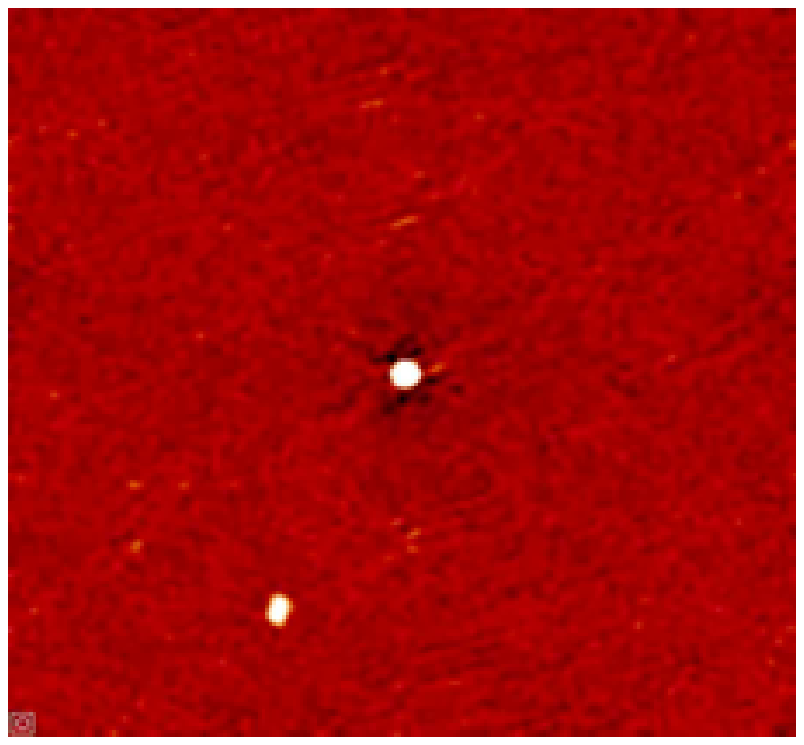}
}
\centerline{
   \includegraphics[height=2.5in]{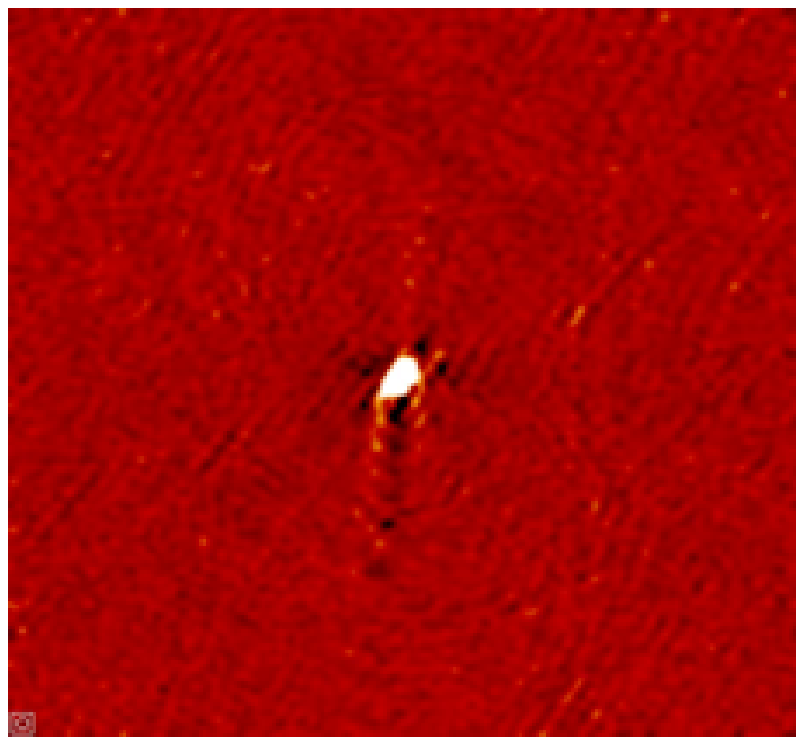}
   \includegraphics[height=2.5in]{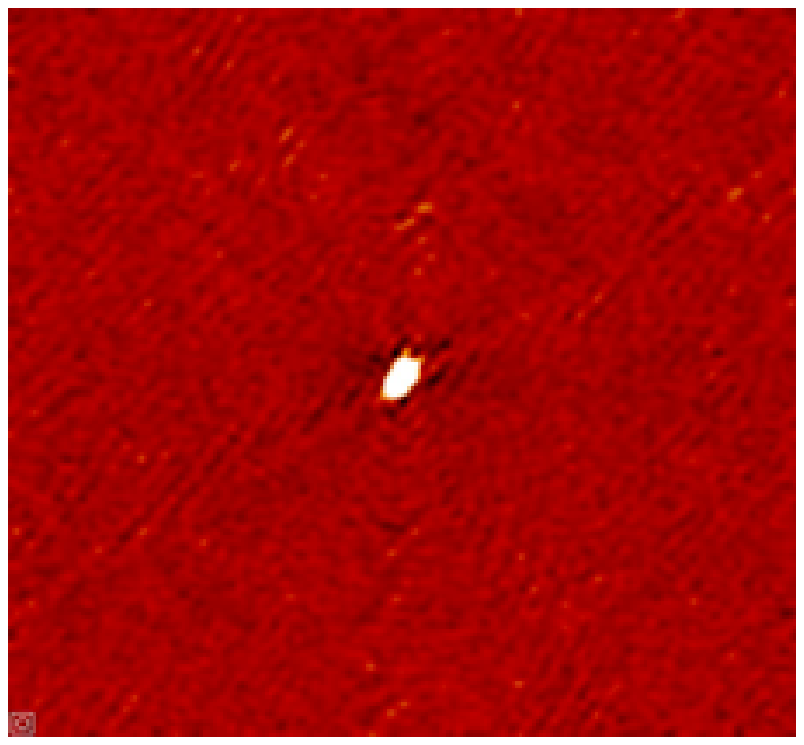}
}
\caption{Ionospheric corrections: Images uncorrected (left column) and the corresponding corrected ones (right column).  The rms noise levels are 0.30~mJy/beam (uncorrected images) and 0.25~mJy/beam (corrected images).  The sources have peak fluxes of 210~mJy/beam (upper set) and 356~mJy/beam (lower set) and are dynamic-range limited even after the ionospheric correction although the residuals in the corrected images are reduced significantly.}
\label{image-pairs}
\end{figure*}

\begin{figure*}
\centering
\centerline{
   \includegraphics[height=2.7in]{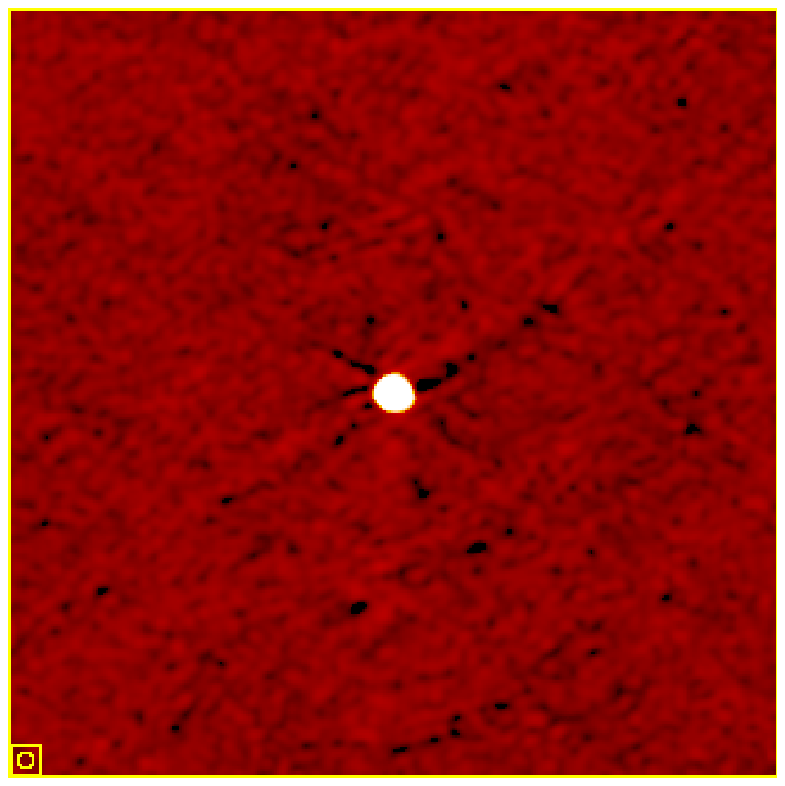}
   \includegraphics[height=2.7in]{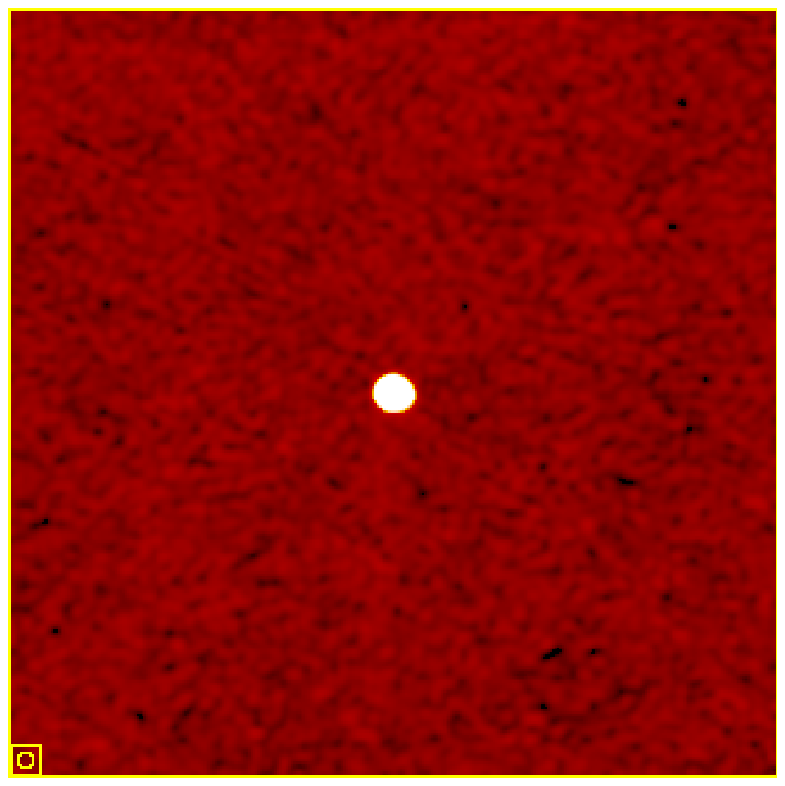}
}
\centerline{
   \includegraphics[height=2.85in]{histo-source-nopeel.eps}
   \includegraphics[height=2.85in]{histo-source-peel.eps}
}
\caption{Peeling improves the images and removes non-statistical errors as evidenced by the comparison of the images of the 2nd strongest source (observed peak of 310~mJy/beam) uncorrected (left column) and the corresponding corrected one (right column).  The histograms have been obtained using a region with an area of $\sim$$1000$~synthesized beams.  The rms noise levels are 0.35~mJy/beam (unpeeled) and 0.31~mJy/beam (peeled), still larger than over the full, peeled image (0.25~mJy/beam) indicating low-level, residual distortions even after the peeling self-calibration.}
\label{peeled-images}
\end{figure*}

\section{Peeling}
\label{peel}

The improvements from the ionospheric corrections are quite noticeable but the brightest sources are dynamic-range limited and artifacts are still visible with sidelobes that are far-reaching as usual.  Due to the chromaticity of interferometric arrays, such artifacts can affect even spectral images as the error pattern will scale with frequency, and the superposition of the sidelobes from such error patterns will vary slowly with frequency which can generate spurious spectral signals at {\it a priori} random positions.  These errors are likely due to a variety of reasons such as higher-order distortions in the ionosphere (turbulence) and rapid variability (corresponding to time-scales shorter than 2-minutes in our example) that cannot be corrected due to limited sensitivity, and also to departures from the actual primary-beam response of the antennas from the average azimuthally-symmetric function that is implicitly assumed if no time-variability of the beam is allowed for (given the alt-az antenna mounts at the VLA).  Limited sensitivity precludes correction for any short-term variability but we have attempted to correct for the (presumably slow) directionally-dependent variations in the response of the primary beam using ``limited peeling.''  This is a variation of the scheme proposed by Noordam for the calibration of the LOFAR array [16] which we discuss next.

Peeling is essentially a partial self-calibration of the brightest sources in order to reduce their associated error patterns which proceeds sequentially.  It is a non-linear operation which can generate ghosts as it is difficult to modify the calibration of a small region of the sky without affecting the images of other sources.  Indeed, the (local) self-calibration zeroes the (residual) sidelobes of any far sources at the position of the source being ``peeled'' which in turn affects such sources.  The Obit implementation proceeds differently by imaging, deconvolving and subtracting all other sources prior to performing the peeling self-calibration on the source being processed in order to minimize the impact of the self-calibration on positions outside the small field over which it will be applied.

Following the imaging steps described in the previous section, the models of all the fields except the one to be peeled are subtracted from the data using the field-based calibration. This should remove most of the effect of other sources in the FOV such that the residual data set is dominated by the source (field) to be peeled. This residual data set is then imaged, phase selfÐcalibrated, and finally amplitude-and-phase self-calibrated to arrive at the best model of this source. The inverse of the peeling self-calibration solution is then applied to the improved model of this field and subtracted from the original visibilities.  In other words, the best model for the field is ``degraded'' to the calibration state of the full data set and subtracted from it.  This avoids performing/undoing sets of self-calibration operations to the data which minimizes instabilities in the determination and application of the peeling solutions.  The residual visibilities now represent all sources except for the one just ``peeled'' which has been effectively eliminated from the data.  Normal field-based imaging and deconvolution proceeds with iteration of the peeling procedure until no facets remain with peak brightness above the peeling threshold.  Thus, the ``peeled'' sources are eliminated from the data in sequence while minimizing their effect on the FOV.  When the deconvolution of the remaining, non-peeled sources is complete, the CLEAN components from the peeled models are copied to the corresponding table on the facet from which each peeled source was removed. Thus, when the components are restored to the images and the final image is flattened, the models used to peel the strong sources are included.  Details will be given elsewhere but the results can be seen in figure~\ref{peeled-images}.  Although the rms noise over the full image is lowered only by a small amount, the rms noise over the immediate field to a bright source is lowered by about 10\% and the distribution of residuals is also greatly improved as shown in the histogram which is determined from the pixels within an area surrounding the source covering $\sim$1000 synthesized beams.  The final image of this nominally ``blank field'' contains a great many sources (Fig. 6) with a total flux of 18.9~Jy whereas the discarded ``flanking fields'' totaled 1.4~Jy (both totals without correction for the attenuation of the primary beam).

\begin{figure}
   \centering
   \includegraphics[height=3.5in]{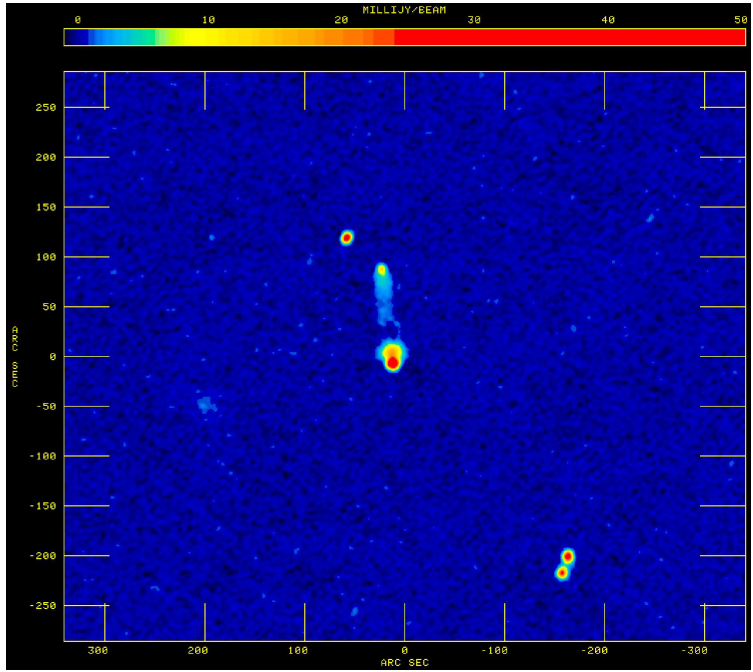}
\caption{ 
A small area of the corrected image with a resolution of $6''$ and an rms noise level of 0.25~mJy/beam.  This sub-image is centered at J130057+300656 and the tick-mark separations are of $100''$ (horizontal axis) and $50''$ (vertical axis).  The color scale is logarithmic, from $-1$~mJy/beam to $+50$~mJy/beam.  Although chosen to be devoid of strong sources and thus as ``empty'' as could be found, clearly there is no such thing as an ``empty field'' at low frequencies given the large FOV seen by the VLA antennas.
}
\label{detail}
\end{figure}

\section{Discussion}
\label{disc}

Despite the obvious improvement brought about by the ionospheric correction, the dynamic range is still limited by artifacts on strong sources.  We have investigated its possible cause and find that it is not due to a problem with the ionospheric correction as it does not depend on the residual seeing threshold discussed above.  It is most likely due to poor knowledge of the primary beam response. 

Indeed, the physical differences amongst the antennas lead to differences in their radiation patterns which depart from the simple, theoretical description in different ways which cannot always be measured with sufficient precision.  In order to make images with high dynamic range it is necessary to derive position- and time-dependent corrections that are different for each antenna. For example, changes in the elevation of the antennas will induce deformations of the back-structure and primary beam that are different for each antenna and that cannot be predicted theoretically with sufficient accuracy at the present time.  In addition, the antennas are out-of-focus by different amounts with a peak-to-peak range of $\sim$$10$~cm due to differences in the relative location of the primary reflector with respect to the feed legs (which are corrected by the focusing mechanism for the Cassegrain frequencies but cannot be equalized at this band due to mechanical constraints [11,12]).  We have addressed this problem with limited peeling of the strong sources and find some improvement using moderate time intervals, significantly longer than those that are used to evaluate and correct the ionospheric ``seeing.''

The method allows for a ÒcontrolledÓ change in the response of the antennas at the position of a few strong sources after the determination of the images of the weaker sources which are removed from the measured visibilities and restored after the ÒpeelingÓ has converged.  The peeling is essentially an independent self-calibration of the strongest sources on timescales that are sufficiently long to achieve convergence, thus enabling only the correction of slowly varying gains (such as the first-order changes arising from the departure of azimuthal symmetry in the beam response discussed above).  Our method modifies the technique proposed by Jan Noordam [16] in the subtraction from the visibilities of the best representation of all of the sources except the one being peeled at each step in order to avoid the propagation of ``ghost'' images.  In addition, convergence is better achieved by subtracting the (corrupted) peeled sources from the original (i.e. previous to the peeling operation) visibilities.  This is achieved by undoing the self-calibration on the clean components that describe the peeled source and subtracting those from the visibilities rather than calibrating/uncalibrating the visibilities at each peeling step.  The procedure operates on several sources in succession.  Although it is possible to iterate the peeling on the full set of bright sources, degrees of freedom are used up quickly.

It is obvious that the procedure works best on strong sources but one must beware of the noise bias.  The procedure appears to work on suitably long time-scales.  However, it is hard to obtain convergence on intermediate sources  and on short time-scales which set the ultimate limit to the dynamic range that can be achieved.

\section{Conclusions}
\label{conc}

We have shown that under reasonable observing conditions the quality of images derived from the VLA observations at $\sim$327~MHz is limited by ionospheric effects that can be corrected to a large extent by adapting the ``field-based'' calibration technique used at lower frequencies of modeling the ionosphere using a basis set of low-order Zernike polynomials.  The images show a large number of truly point sources as they are described by single-pixel Ôclean components.Õ  However, the images have limited (local)  dynamic range as artifacts appear in the vicinity of some of the strongest sources, most likely due to imperfect knowledge of the variable and non-azymuthally-symmetric antenna primary beam.  The use of limited ``peeling'' leads to improved images but is limited by the necessary sensitivity on short time-scales.  Efforts are under way to improve this correction which should benefit from larger bandwidths if the radio-frequency interference can be overcome, which will likely transfer the problem to yet some other systematic error, possibly the variability of the bandpass response which would have to be addressed with an algorithm to perform some sort of  ``bandpass self-calibration.''




\section*{Acknowledgements}
The data used in this paper correspond to an observation of the ``Mitchell-Condon'' field by our collaborator Ken Chambers which will be published in detail elsewhere.  We thank Sylvie Cabrit and Thibaut LeBertre for their questions and suggestions that have helped us to improve this text.  We have used the Obit package (http://www.cv.nrao.edu/$\sim$bcotton/Obit.html) and the NVSS catalog [13].

\end{document}